\documentstyle[preprint,aps]{revtex}

\begin{document}


\title{Stability and collapse of a hybrid Bose-Einstein condensate of
atoms 
and molecules}

\author{Sadhan K Adhikari}

\address{Instituto de F\'{\i}sica Te\'orica, Universidade Estadual
Paulista, 01.405-900 S\~ao Paulo, S\~ao Paulo, Brazil}

\date{\today}
\maketitle
\begin{abstract}

The dynamics of stability and collapse of a trapped atomic Bose-Einstein
condensate (BEC) coupled to a molecular one is studied using the
time-dependent Gross-Pitaevskii (GP) equation including a nonlinear
interaction term which can transform two atoms into a molecule and vice
versa. We find interesting oscillation of the number of atoms and
molecules for a BEC of fixed mass.  This oscillation is a consequence of
continuous transformation in the condensate of two atoms into a molecule
and vice versa. For the study of collapse an absorptive contact
interaction and an imaginary quartic three-body recombination term are
included in the GP equation. It is possible to have a collapse of one or
both components when one or more of the nonlinear terms in the GP equation
are attractive in nature, respectively.

{\bf PACS Number(s): 03.75.Fi, 05.30.Jp}

\end{abstract}

\newpage
\section{Introduction}

The experimental detection \cite{1,1x,1a} of Bose-Einstein condensation
(BEC) 
at ultralow temperature in dilute trapped alkali metal and
hydrogen atoms  has initiated intense theoretical activities in this
subject \cite{2,2x,2a,2b,2c,2d,2e,rex,3,4,6}. Many properties of the
condensate
are
usually described by the nonlinear meanfield Gross-Pitaevskii (GP)
equation
\cite{8} which can be considered to be a generalization of the
Schr\"odinger equation in the presence of interatomic interaction leading
to a nonlinear term.  

Two interesting features of BEC are (a) the collapse in the case of an
attractive atomic interaction \cite{1a,2} and (b) the possibility of the
formation of a hybrid atomic and molecular condensate
\cite{2a,2b,2c,2d,2e}
coupled
by coherent Raman transition \cite{ram} or Feshbach resonance
\cite{2c,fesh}.
Although there has been experimental observation of a coupled BEC
involving two atomic states \cite{cp}, so far there has been no
experimental
realization of a molecular BEC.  Because of the intrinsic interest in a
hybrid
atomic and molecular BEC, in this work the dynamics of  such a  BEC is
studied using the time-dependent GP equation
with special attention to its stability and collapse.
 For the case of a trapped
 two-species BEC a large repulsive interaction between the two
species can induce instability in the system \cite{re}.

For attractive interatomic interaction \cite{1a,2}, the condensate is
stable for a maximum critical number of atoms.  When the number of atoms
increases beyond the critical number, due to interatomic attraction the
radius of BEC tends to zero and the central density tends to infinity.
Consequently, the condensate collapses emitting atoms until the number of
atoms is reduced below the critical number and a stable configuration is
reached. With a supply of atoms from an external source the condensate can
grow again and thus a series of collapses can take place, which was
observed experimentally in the BEC of $^7$Li with attractive interaction
\cite{1a,2}. Theoretical analyses based on the GP equation also confirm
the collapse \cite{1a,2,6}.

The possibility of the formation of a hybrid trapped atomic BEC coupled to
a BEC of diatomic molecules via either Raman photoassociation \cite{ram}
or a Feshbach resonance \cite{2c,fesh} has been suggested and studied
recently \cite{2a,2b}. These authors used a nonlinear interaction term
\cite{2a,lee} in the Hamiltonian that allows for a coherent
number-conserving nondissipative transition of two atoms to a molecule and
vice versa. Possible candidates for molecular BEC are $^{87}$Rb$_2$,
$^{85}$Rb$_2$, $^{23}$Na$_2$ etc \cite{2b,t}.  The coherently coupled BEC
could allow localized solitons even in the absence of a trap \cite{2a}.
However, in this work we shall only consider the interesting dynamics of
this hybrid BEC in the presence of a trap. Because of the general interest
there have been several theoretical studies involving two coupled atomic
condensates \cite{xx}

Here we critically study the atomic plus molecular hybrid BEC in the
presence of attractive bosonic interaction(s). The usual GP equation
allowing for transition of two atoms to a molecule and vice versa
conserves the overall number of atoms. In the study of the
dynamics of this BEC, oscillation of the individual
numbers of atoms and molecules are found with the conservation of overall
total number (number of atoms plus twice the number of diatomic
molecules). The dynamics of collapse for attractive interaction is studied
by introducing 
absorptive contact interactions which allow for a growth in the particle
numbers from an external source in addition to  imaginary quartic
three-body interactions responsible for recombination loss \cite{2}. In
the
presence of the imaginary contact  and three-body interactions
the GP
equation does not conserve the overall number of atoms.  If the strengths
(and signs) of these two imaginary interactions are properly chosen, the
solution of the GP equation could produce a growth of the condensate with
time for numbers of atoms and molecules less than the respective critical
numbers. Once they increase past these critical numbers the three-body
interaction takes control and these numbers suddenly drop below the
critical level by recombination loss signaling a collapse \cite{2}. Then
the absorptive term takes over and the individual numbers start to
increase. This continues indefinitely showing an infinite sequence of
collapses.

There are three types of nonlinear interactions in the GP equation for the
hybrid BEC of atoms and molecules:  e.g., between two atoms, between two
molecules, and between an atom and a molecule. If one of these is
attractive, one component of the condensate may experience collapse. If
two are attractive one can have collapse in both components. Specifically,
one can also have collapse of both components if only the interaction
between an atom and a molecule is sufficiently attractive.

In Sec. II we present a set of coupled GP equations which is appropriate
for the study of the hybrid atomic-molecular BEC including a new
nonlinear interaction term which allows for the possibility of
transformation of two  atoms in the condensate to a molecule and vice
versa. In addition we include an absorptive contact term and a three-body
recombination term appropriate for studying collapse of the BEC. In Sec.
III we solve the coupled GP equation numerically and study the hybrid
atomic-molecular BEC under different possibilities of atomic and molecular
interactions. Finally, in Sec. IV we give some concluding remarks.

\section{Coupled Gross-Pitaevskii Equation for Atomic-Molecular
Condensate}

Here, in addition to the kinetic energies, trap
potentials, and usual nonlinear interaction terms   the following
interaction  responsible for a nondissipative  transfer of two atoms into
a
molecule and vice versa is also considered in the hamiltonian
\cite{2a,lee}
\begin{equation}\label{1}
H_{\mbox{int}}= \frac{\chi}{2}\int d {\bf r}\left[ \hat
\psi_1^2 \hat
\psi_2
^\dagger 
+ \hat \psi_1 ^{\dagger 2} \hat \psi_2   \right],
\end{equation}
where $\hat \psi_ 1$ ($\hat \psi _ 2$) is the complex atomic (diatomic
molecular)
field and $\chi$  the coupling. For  a nonzero $\chi$ neither the number
of atoms nor that of the molecules is conserved. However the total mass of
all the atoms and molecules is conserved. In a BEC the chemical conversion
of two atoms to a diatomic molecule is dominated by coherent stimulated
emission, in which transitions are enhanced by the number of molecules
already occupying the ground state. The conversion of atoms to
molecules could be
resonant via a Feshbach resonance. Or, it could be effected via stimulated
Raman coupling introduced by two laser fields. In this fashion it is
possible to select a specific molecular state in the condensate. 

In the presence of the extra interaction (\ref{1}) 
 the mean-field GP equation for wave-function components  become
\cite{2a,2b}
\begin{eqnarray}\label{cc} \biggr[
&-&\frac{\hbar^2}{2m }\frac{1}{r}\frac{\partial^2 }{\partial
r^2}r + \frac{1}{2}m\omega^2 r^2 +\sum_{j=1}^2
g_{1j}|\psi_j({
r},\tau)|^2 \nonumber 
\\
&-& \mbox{i}\hbar\frac{\partial}{\partial \tau}\biggr]
\psi_1(r,\tau)+\chi \psi_2(r,\tau) \psi_1^*(r,\tau) =0,
\end{eqnarray}
\begin{eqnarray}\label{cd} \biggr[
&-&\frac{\hbar^2}{4m }\frac{1}{r}\frac{\partial^2 }{\partial
r^2}r + c m\omega^2 r^2 +\sum_{j=1}^2
g_{2j}|\psi_j({
r},\tau)|^2 \nonumber \\
&-& \mbox{i}\hbar\frac{\partial}{\partial \tau}\biggr]
\psi_2(r,\tau)+\frac{\chi}{2} \psi_1^2(r,\tau)  =0,
\end{eqnarray}
where $m$ is the atomic mass, $2m$ the diatomic molecular mass,
$g_{ij}$ the nonlinear interaction,  
$\omega$ the
frequency of the harmonic oscillator trap, and $i=1$
(2) denotes an atom (molecule).  The parameter $c$ has been
introduced in  (\ref{cd}) to modify the frequency of the trap for
molecules. The diagonal interaction term $g_{ii}$ is related to the
scattering length $a_i$ of $i$-type bosons via $g_{ii}=4\pi
\hbar^2a_i/m_{i}$, where $m_i$ is the corresponding mass. The
wave-function components are normalized as
\begin{eqnarray}\label{no1}
\int d^3 r |\psi_1(r,\tau)|^2 &= &{N}_1,  \\
\int d^3 r |\psi_2(r,\tau)|^2 &= &{N}_2,
\end{eqnarray}
where $N_j$ is the number of $j$-type condensed boson, 
For $\chi=0 $ the
number of bosons of each type is 
conserved.   
For $\chi \ne 0$, the number of bosons of each type is not separately
conserved
 and 
there will be continuous transfer of one type of bosons to
another and vice versa.  
However,
the overall total number $(N_1+2N_2)$
is conserved.

As in Refs. \cite{4} it is convenient to use dimensionless variables
defined by $x = \sqrt 2 r/a_{\mbox{ho}}$,  $t=\tau \omega, $  and
$\phi_i(x,t) = x\psi_i(r,\tau ) (\sqrt 2\pi a_{\mbox{ho}}^3)^{1/2}$  
where
$a_{\mbox{ho}}\equiv \sqrt {\hbar/(m\omega)}$.
In
terms of these,   (\ref{cc}) and (\ref{cd})  become 
\begin{eqnarray}\label{e}
\biggr[ &-&\frac{\partial^2 }{\partial
x^2} + \frac{ x^2}{4} +\sum_{j=1}^2 n_{1j}
\frac{|\phi_j({x},t)|^2}{x^2}
-\mbox{i}\xi_1\frac{|\phi_1({x},t)|^4}{x^4}\nonumber \\
&+&\mbox{i}\gamma_1
-  \mbox{i}\frac{\partial
}{\partial t} \biggr]\phi_1({ x},t)+ \eta \frac{\phi_2(x,t)
\phi_1^*(x,t)}{x}
=0, \end{eqnarray}
\begin{eqnarray}\label{f}
\biggr[ &-&\frac{1}{2}\frac{\partial^2 }{ \partial
 x^2} + \frac{ c x^2}{2} +\sum_{j=1}^2 n_{2j}
\frac{|\phi_j({x},t)|^2}{x^2}
-\mbox{i}\xi_2\frac{|\phi_2({x},t)|^4}{x^4}\nonumber \\
&+&\mbox{i}\gamma_2
-  \mbox{i}\frac{\partial
}{\partial t} \biggr]\phi_2({ x},t)+ \frac{\eta}{2}
\frac{\phi_1^2(x,t)}{x}
=0, \end{eqnarray}
where  $n_{ij}\equiv  g_{ij}m/(\sqrt 2 \pi \hbar^2 
a_{\mbox{ho}})$ is negative (positive) when the
corresponding interaction 
is attractive (repulsive) and the dimensionless coupling 
$\eta =\chi / [\hbar \omega (\sqrt 2 \pi a^3_{\mbox {ho}})^{1/2} ]$. 
The diagonal elements  of   $n_{ij}$ are $n_{ii}= 2\sqrt 2
(a_i/a_{\mbox{ho}})$.
In  (\ref{e}) and (\ref{f})
a diagonal absorptive (strength $\gamma_i$) and quartic three-body
(strength $\xi_i$) interactions, 
appropriate to study
collapse, are introduced \cite{2}. A nonzero $\gamma_i$ allows the
possibility of the absorption of atoms and molecules from the surrounding
into the
condensate, whereas a nonzero $\xi_i$ allows for the possibility of
ejection of particles from the condensate due to three-body recombination.
Consequently, the total mass of
the condensate is not conserved for nonzero $\gamma_i$ and $\xi_i$.  A
nonzero $\gamma_i$ and $\xi_i$ is appropriate for the study of collapse of
the system.

The fluctuation of atomic and molecular
numbers is best studied via the
quantities
\begin{eqnarray}
 N_{\mbox{at}}=\frac {1}{N_1^{(0)}}\int_0^\infty	|\phi_1(x,t)|^2
dx, \\
N_{\mbox{mol}}= \frac {1}{N_2^{(0)}}\int_0^\infty	|\phi_2(x,t)|^2
dx, 
\end{eqnarray}
where ${N_1^{(0)}}$ and ${N_2^{(0)}}$ are the initial numbers of the two
types.
In the presence of a general absorptive
interaction $\eta\ne 0, \gamma_i\ne 0$ and $\xi_i\ne 0$, 
the quantities $N_{\mbox{at}}$ and
$N_{\mbox{mol}}$  carry the information about time-evolution of the number
of atoms and
molecules
in the condensate.  
For
$\gamma_i=\xi_i=\eta = 0$, the number of atoms and molecules are
separately conserved and one has
$N_{\mbox{at}}=N_{\mbox{mol}}=1.$
For
$\gamma_i=\xi_i=0$ and $\eta \ne 0$, there is no separate number
conservation  for atoms and molecules: 
$N_{\mbox{at}}\ne 1,$
$N_{\mbox{at}}\ne 1;$ although the total mass of the condensate is
conserved. 
 For $\gamma_i\ne 0$ and/or
$\xi_i\ne 0$
there is no
conservation of the total mass of the condensate.

The root-mean-square (rms)
radius   $x_{\mbox{rms}}(t)$  of component $i$ is
given by \begin{equation}\label{7}
x_{\mbox{rms}}(t)=
\left[\frac {\int_0
^\infty x^2 |\phi_i(x,t)| ^2 dx} {\int_0
^\infty  |\phi_i(x,t)| ^2 dx
}\right]^{1/2}.  \end{equation}
The oscillation  of this rms radius denote  radial oscillation of the
condensate.

\section{Numerical Method}

To solve  (\ref{e}) and (\ref{f}) numerically one needs the proper
boundary
conditions as $x\to 0$ and $\infty$. For a confined condensate, for a
sufficiently large $x$, $\phi_i(x,t)$ must vanish asymptotically. Hence
all the nonlinear terms proportional to the product of wave-function
components  can eventually be
neglected in the GP equation for large $x$. Consequently the the
asymptotic form of the physically acceptable solution is given by $
\lim_{x \to \infty} |\phi_i(x,t)|\sim \exp (-{x^2}/{4}).$ Next we consider
 (\ref{e}) and (\ref{f}) as $x\to 0$. The nonlinear terms approach a
constant in
this limit because of the regularity of the wave function at $x=0$ and 
one has the condition $ |\phi_i(0,t)|=0.$

Next we discretize  (\ref{e}) and (\ref{f}) in both space and time by
using a space step $h$ and time step $\Delta $ with a finite difference
scheme using the unknown ${(\phi_i)}^k_p$ which will be approximation of
the exact solution $\phi_i(x_p,t_k)$ where $x_p= p h$ and $t_k=k\Delta $. 
After discretization we obtain a set of algebraic equations which could
then be solved by using the known asymptotic boundary conditions.  The
numerical procedure is similar to that in 
refs. \cite{2,4}.  The time
derivative in these equations involves the wave functions at times $t_k$
and $t_{k+1}$. As in the uncoupled case we express the wave functions and
their derivatives in  (\ref{e}) and (\ref{f}) in terms of the average
over times $t_k$ and $t_{k+1}$ \cite{koo} and the resultant scheme leads
to accurate results and good convergence. In practice we use the following
Crank-Nicholson-type scheme \cite{koo} to discretize the partial
differential equations (\ref{e}) and (\ref{f}) 
 \begin{eqnarray} &i&\frac{(\phi_1)_p^{k+1}-(\phi_1)_p^{k} }{\Delta } =
-\frac{1}{2h ^2}\biggr[(\phi_1)^{k+1}_{p+1}-2 (\phi_1)^{k+1}_{p}\nonumber
\\&+&(\phi_1)^{k+1}_{p-1} + (\phi_1)^{k}_{p+1}-2
(\phi_1)^{k}_{p}+(\phi_1)^{k}_{p-1}\biggr]\nonumber \\
&+&\frac{1}{2}\left[\frac{x_p^2}{4}+\sum_{l=1}^2
n_{1l}\frac{|(\phi_l)_p^{k}|^2}{x_p^2}-i\xi_1
\frac{|(\phi_1)_p^{k}|^4}{x_p^4}+i\gamma_1 \right]\nonumber \\ &\times&
\left[(\phi_1)_p^{k+1}+(\phi_1)_p^k\right]+\frac{\eta(\phi_2)_p^k}{x_p}
{(\phi_1)_p^k}^* , \label{fx} \end{eqnarray}
 \begin{eqnarray} &i&\frac{(\phi_2)_p^{k+1}-(\phi_2)_p^{k} }{\Delta } =
-\frac{1}{4h ^2}\biggr[(\phi_2)^{k+1}_{p+1}-2 (\phi_2)^{k+1}_{p}\nonumber
\\&+&(\phi_2)^{k+1}_{p-1} + (\phi_2)^{k}_{p+1}-2
(\phi_2)^{k}_{p}+(\phi_2)^{k}_{p-1}\biggr]\nonumber \\
&+&\frac{1}{2}\left[\frac{cx_p^2}{2}+\sum_{l=1}^2
n_{2l}\frac{|(\phi_l)_p^{k}|^2}{x_p^2} -i\xi_2
\frac{|(\phi_2)_p^{k}|^4}{x_p^4}+i\gamma_2 \right]\nonumber \\ &\times&
\left[(\phi_2)_p^{k+1}+(\phi_2)_p^k\right]+\frac{\eta 
(\phi_1)_p^k}{2x_p}
{(\phi_1)_p^k}. \label{fx2} \end{eqnarray}

Considering that
the wave function components $\phi_i$  are  known at time $t_k$, 
(\ref{fx}) and (\ref{fx2}) are   equations in
the unknowns $-$ $(\phi_i)_{p+1}^{k+1},(\phi_i)_p^{k+1}$ and
$(\phi_i)_{p-1}^{k+1}$.
In a lattice of $N$ points  (\ref{fx}) and (\ref{fx2}) represent a
tridiagonal set for
$p=2,3,...,(N-1)$ for a specific component $\phi_i$. This set has a unique
solution if the wave functions at
the two end points $(\phi_i)_{1}^{k+1}$ and $(\phi_i)_{N}^{k+1}$ are
known \cite{koo}. In the
present problem these values at the end points are provided by the known
asymptotic boundary conditions. This tridiagonal set of equations is
solved by Gaussian elimination and back-substitution \cite{koo}. 

To solve  (\ref{fx}) and (\ref{fx2}) we employ space step $h=$ 0.0001
with $x_{\mbox{max}}\le 15$ and time step $\Delta$ =0.05 with
$t_{\mbox{max}}=100$. After some experimentation we find that these values
of the steps give good convergence. The iteration is started with the
known normalized (harmonic oscillator) solution of  (\ref{e}) and
(\ref{f}) obtained with $n_{ij}=\gamma_i=\xi_i=\eta=0$ with
normalization
\begin{equation}
\int _0 ^\infty |\phi_i(x,t)|^2= N_i^{(0)}, \quad i=1,2,
\end{equation}  where $N_i^{(0)}$ is the initial number of atoms.
The nonlinear
parameters $n_{ij}$ and $\eta$ are increased by equal amounts over 500 to
1000 time iterations starting from zero until the desired final values are
reached.  During this process we keep $\xi_i=\gamma_i=0$.  The resulting
solution is the ground state of the condensate corresponding to the
specific nonlinearity.

The time-dependent approach is the most suitable for solving
time-evolution problems. In the present study we only consider evolution
problems starting from a stable condensate at $t=0$. In these cases the
stationary problem is solved first and the wave function so obtained
serves as the starting wave function for the time-evolution problem.

To study the effect of an external source of bosons on preformed
condensate(s)  we need to set $\gamma_i \ne 0$ and $\xi\ne 0$ and
perform the time iteration of the GP equations (\ref{e}) and (\ref{f})
starting at $t=0$ from the solution obtained with $\gamma_i  =0$ and
$\xi_i  =0$. A nonzero $\gamma_i$ will make the condensate grow in size
absorbing bosons from outside. In the case of attractive interaction(s),
as the condensate grows past the critical size a nonzero $\xi$ will allow
for three-body recombination so that the condensate can shrink and
experience the so called collapse by emitting particles. Hence a nonzero
$\gamma_i$ and $ \xi_i$ could simulate the dynamical growth of the
condensate in general as well as its collapse for attractive interaction.

\section{Numerical Result} 

Instead of concentrating on a specific  atomic-molecular BEC,
here we  study some universal features of such a
BEC related to stability and collapse. For a specific application we
require the knowledge of the different scattering lengths: those  for two 
atoms, two molecules and between an atom and a molecule. At present these
scattering lengths are not all known for a specific system.

\subsection{Repulsive Interaction: Stable Condensate}

First we consider the nondissipative solutions of   (\ref{e}) and
(\ref{f})
obtained with $\eta\ne 0$ and  $\gamma_i=\xi_i=0$ for all repulsive
interactions and
demonstrate the oscillation of numbers of atoms and molecules. In this
case the condensate increases in size indefinitely with time if a nonzero
$\gamma_i$ and $\xi_i$ are used. This is why it can be termed stable.
However, there is spontaneous transformation of atoms to molecules and
vice versa even for $\gamma_i=\xi_i=0$ and the number of atoms and
molecules oscillate with time. The numerical results are convergent for
small values of nonlinear parameters $n_{ij}$ and $\eta$. Here to
demonstrate the viability of the present numerical scheme we present
results for reasonably large values of $n_{ij}$.  Specifically,
we consider $c=1/16$, $N_1^{(0)}=N_2^{(0)}=10000$, $n_{11}=n_{22}=0.001$,
$n_{12}=n_{21}=0.0005,
x_{\mbox{max}}=15$ and
$\eta=0.001, 0.005,$ and $ 0.01$.  
For a typical experimental situation \cite{rex} this corresponds to 
$2\sqrt{2}|a_i|/a_{\mbox{ho}} =10^{-3}$.
The solution of the complete  coupled GP equations
(\ref{e}) and (\ref{f})
is obtained
by iteration starting with the solution of the uncoupled
Schr\"odinger equation for the harmonic oscillator problem obtained 
by setting $n_{ij}=\eta=0$ in these equations. The nonlinear
parameters $n_{ij}$  are increased in small steps during
iteration until their final values are reached after 1000 iterations. 
We then obtain the desired
solution of the coupled GP equation. 
In figure 1 (a)  we plot the atomic and
molecular wave functions $\phi_1$ and $\phi_2$ obtained from this
procedure. 

We continue the time iteration 
after the solutions of the coupled GP equation shown
in figure 1 (a) are  obtained at a specific time, which we now call  
$t=0$, without
changing the nonlinear parameters
anymore. In figure 1 (b) we plot the quantities   $ N_{\mbox{at}}$ and
$ N_{\mbox{mol}}$ vs. time $t$ for $t>0$.
For $\eta\ne 0$, as shown in figure 1 (b), the numbers $ N_{\mbox{at}}$
and
$N_{\mbox{mol}}$ undergo rapid oscillations.  
In figure 1 (c) we plot the
corresponding rms
radii of the atomic
and molecular condensates for $t>0$.  We find that the rms radii also
execute
oscillation along with $N_{\mbox{at}}$ and $N_{\mbox{mol}}$ due to
continued dynamical conversion of atoms to molecules and vice versa. The
oscillation of the rms radii is related to the radial oscillation of the
condensate. 

\subsection{Attractive Interaction: Route to Collapse}

Now we consider the case of attractive atomic and molecular
interaction(s) which may lead to collapse. We note that  $\eta$
is a nonlinear parameter in the coupled GP equation and a large value of
the nonlinear parameter 
$\eta$ will imply  a large
numerical error in the solution. A very small  $\eta$ on the other hand
will not give adequate feedback between atoms and molecules.  
We study collapse in the hybrid BEC under several conditions with
$\eta =0.00625 $, as this value of $\eta$ gives
adequate feedback between
atoms and molecules and accurate numerical result. 

First we  study  the simplest case of
collapse by taking all interactions 
to be attractive 
corresponding to a  negative $n_{ij}$ for all $i$ and $j$.
Now 
both components of BEC could experience collapse. 
In this case
the wave function components calculated numerically as 
in figure 1 (a) at $t=0$  are plotted in figure 2 (a) for
$N_1^{(0)}=N_2^{(0)}=
1600,   n_{11}=-0.000975,
          n_{22}=-0.000475, 
          n_{12}= n_{21}=-0.000125, $
 $c=1/16$, $\eta=0.00625$, $x_{\mbox{max}}=15$, with
$\xi_i=\gamma_i=0$.  
In
the case of attractive interaction, for BEC with $^7$Li the typical number
of atoms in the condensate is 10$^3$ and
$|a_i|/a_{\mbox{ho}} \sim
10^{-3}$ to 10$^{-4}$ \cite{1a}. In this example, $|a_1|/a_{\mbox{ho}}\simeq
3.5\times 10^{-4}$, 
and $|a_2|/a_{\mbox{ho}}\simeq 1.7 \times 10^{-4}$.
Hence the present numbers
are well within the experimental scenario of BEC for atoms with attractive
interaction.  From figure 2 (a) we find that the wave-function components
are more centrally peaked compared to those in figure 1 (a). This
corresponds to small rms radii and large central density denoting an
approximation to collapse.

Although the collapse of the coupled condensates could be inferred from
the wave functions of figure 2 (a), we also study the dynamics of the
collapse from a time evolution of GP equations (\ref{e}) and (\ref{f}) for
$t>0$ in the presence of an absorption and three-body recombination, e.g.,
for $\gamma_i \ne 0$ and $\xi_i \ne 0$ \cite{2}. We consider the solution
of   (\ref{e}) and (\ref{f})  as shown in figure 2 (a) at $t=0$ and
allow this solution to evolve in time with $\gamma_i \ne 0$ and $\xi_i \ne
0$ by iteration. The general nature of time evolution is independent of
the actual values of $\gamma_i$ and $\xi_i$ employed, provided that a very 
small value for $\xi_i (\sim 10^{-6})$ and a relatively larger one for
$\gamma_i (\sim 0.1$)  are chosen \cite{2}.  In actual calculation we
took $\gamma_1=0.15, \gamma_2=0.08 $ and $\xi_1=  0.00000031,
\xi_2=0.00000094$. The quantities 
 $N_{\mbox{at}}$ and $N_{\mbox{mol}}$ show the succession of
collapses in a transparent fashion and we plot them in figure 2 (b) vs.
time
$t$. These quantities  increase with time due to the absorptive
interaction
($\gamma_i$). Once they are larger than the respective critical numbers
for collapse,  the three-body recombination takes over and these
quantities
fall below the critical level and the system collapses. Then the
absorptive term is operative and these quantities  increase again
resulting
in a
sequence of collapses.  In figure 2 (c)  we plot the rms radii
$x_{\mbox{rms}}$ of the atomic and molecular condensates of the model of
figure 2 (b) undergoing collapse for $t>0$. The rms radii of the
components
execute oscillation similar to those in the uncoupled case \cite{2}.

The collapse was studied under various situations. After numerical
investigation we find if only the diagonal parameter $n_{11}$ or $n_{22}$
is attractive and all other $n_{ij}$'s are repulsive then it is possible
to have a collapse of component 1 or 2.  If in addition to one diagonal
parameter $n_{ii}$, the nondiagonal parameters $n_{12}$ and $n_{21}$ are
attractive, it is possible to have a collapse in both components. If both
diagonal parameters $n_{11}$ and $n_{22}$ are attractive, one can have
collapse of both the components. 

Instead of reporting all our studies, 
in the following we consider  the special case of collapse in the case
where
the diagonal elements $n_{11}$ and $n_{22}$ are positive and the
nondiagonal terms $n_{12}$ and $n_{21}$ are negative. This corresponds
to repulsion between two atoms and between two molecules and an attraction
between a molecule and an atom. This case is of relevance, as  
 experimentally, the rubidium
system has been measured to have strongly attractive atom-molecule
interaction and repulsive atom-atom and molecule-molecule interactions
\cite{1x,2b,2e}. Also, it is now possible to vary continuously the nature
(attractive or repulsive) and strength of interaction between bosons
\cite{yy1} which can create a specific situation of  atom-molecule,
atom-atom and molecule-molecule interactions.  
We find that when $n_{12}$ and
$n_{21}$ are sufficiently attractive, they can surplus the repulsive terms
$n_{11}$ and $n_{22}$ in  (\ref{e}) and (\ref{f})  and the overall
effective contribution of the quadratic nonlinear terms in these equations
becomes attractive and one can have a collapse of both the components. To
illustrate this interesting effect 
we study numerically  the collapse 
for a negative
$n_{12}$  and $
n_{21}$ and positive $n_{11}$ and $n_{22}$.  In figure 3 (a)  we plot
the
wave-function components obtained by  solving  (\ref{e}) and
(\ref{f})
 with $N_1^{(0)}=N_2^{(0)}=1600,$
$n_{11}= 0.0005,  n_{22}=0.000625,$ 
$n_{12}=n_{21}=-0.00144, $  $c=1/16$, $\eta=0.00625$,
$x_{\mbox{max}}=10$ and
$\gamma_i=\xi_i=0$.
Both wave-function components are  peaked near $x=0$
and have small rms radii. The system collapses with a small increase
in $|n_{12}|$ and/or $|n_{21}|$.
The time evolution of  $N_{\mbox{at}}$ and  $N_{\mbox{mol}}$, 
and the corresponding rms radii are found by solving  (\ref{e}) and
(\ref{f}) iteratively with nonzero $\xi_i$ and $\gamma_i$ 
starting with the wave function  presented in figure 3
(a) at $t=0$. For
$t>0$ we employ $\xi_1= 0.00000031, 
\xi_2= 0.00000075$ and $\gamma_1= 0.12 ,
\gamma_2=0.05$.
The quantities $N_{\mbox{at}}$ and $N_{\mbox{mol}}$ show the
dynamics of collapse directly and they are plotted in figure 3 (b). The
successive growth and decay  of these numbers with time exhibit the
sequence of 
collapses.  Finally, in figure 3 (c) we plot the evolution of the
corresponding rms radii.
The results for $0<t<100$ in figures 2 and 3 are calculated
with
2000 iterations of the GP equation (\ref{e}) and (\ref{f}) using a time
step 0.05.

\section{Conclusion}

To conclude, we studied the stability and collapse of a  trapped hybrid
BEC
consisting of atoms and molecules \cite{2a,2b,2d,2e} coupled by coherent
Raman
transition \cite{ram} or
 Feshbach resonance \cite{2c,fesh}
using the time-dependent coupled 
GP equation. An additional  nonlinear interaction between the atoms and
molecules 
given by  (\ref{1}) is introduced in the formulation  which allows
spontaneous
transformation of
two atoms
into a molecule and vice versa. The effect of an external source of boson
on the condensate is simulated by the absorptive term $\gamma_i$ and the
effect of three-body recombination necessary for the study of collapse is
included via the quartic nonlinear term $\xi_i$ in the GP equation. 

 We find interesting oscillation of the
number of atoms and molecules due to the transformation of atoms to
molecules and vice versa even when all interactions are repulsive.
The component $i$ of the condensate could experience
collapse when the interaction among bosons of type $i$ is attractive.
Both
components could experience collapse when at least the interaction between
a molecule and an atom is attractive.  They could also experience collapse
when two of the bosonic interactions are attractive.  
The time
evolution of collapse is studied via the time-dependent GP equation with
absorption and three-body recombination.  The number of particles of the
component(s)  of BEC experiencing collapse alternately grows and decays
with time. The rms radii of the collapsing condensate also
exhibit interesting
oscillation as in the uncoupled case\cite{2}. The oscillation of the rms
radii denotes a radial oscillation of the condensates which can be
observed experimentally.
Recently, we studied the possibility of collapse in a coupled atomic BEC 
without any interaction allowing for transformation of  one type of atom
to another
\cite{ad}. 
 With the possibility of observation of coupled  BEC, the
findings of this study and of ref. \cite{ad}  could be verified
experimentally in the future.

We thank Dr. Kh. F. Abdullaev for a brief  discussion.
The work is supported in part by the Conselho Nacional de Desenvolvimento
Cient\'\i fico e Tecnol\'ogico  and Funda\c c\~ao de Amparo \`a Pesquisa
do
Estado de S\~ao Paulo of Brazil.

\newpage

{\bf Figure Caption:}

1. (a) Atomic and molecular wave function components $\phi_1(x)$ 
and $\phi_2(x)$
 vs. $x$ at $t=0$, (b) the corresponding numbers $N_{\mbox{at}}$ and
$N_{\mbox{mol}}$
vs. time
$t$ $( > 0)$, and (c) the rms
radii 
vs. time $t$ $(>0) $ obtained 
for  
$N_1^{(0)}=N_2^{(0)}=10000, n_{11}=n_{22}=0.001$,
$n_{12}=n_{21}=0.0005, $  $c=1/16$,
for  
$\xi_i=\gamma_i= 0$.
The  curves are labelled by the corresponding values of
$\eta$.

2. The same as in figure 1 with $N_1^{(0)}=N_2^{(0)}=1600,
n_{11}=-0.000975, n_{22}=-0.000475$, 
$n_{12}=n_{21}=-0.000125, $ $c=1/16$,
and $\eta =0.00625$. In (a) $\xi_i=\gamma_i= 0$, and 
in (b) and (c)  
$\xi_1= 0.00000031, \gamma_1=0.15, 
\xi_2= 0.00000094, \gamma_2=0.08.$

3. The same as in figure 1 with $N_1^{(0)}=N_2^{(0)}=1600,
n_{11}=0.0005, n_{22}=   0.000625$,  
$n_{12}=
n_{21}=-0.00144, $ $c=1/16$,
and $\eta =0.00625$. 
In (a) $\xi_i=\gamma_i=
0$, and in (b) and (c)  $\xi_1= 0.00000031,
 \gamma_1=0.12, \xi_2=
0.00000075, \gamma_2=0.05.$

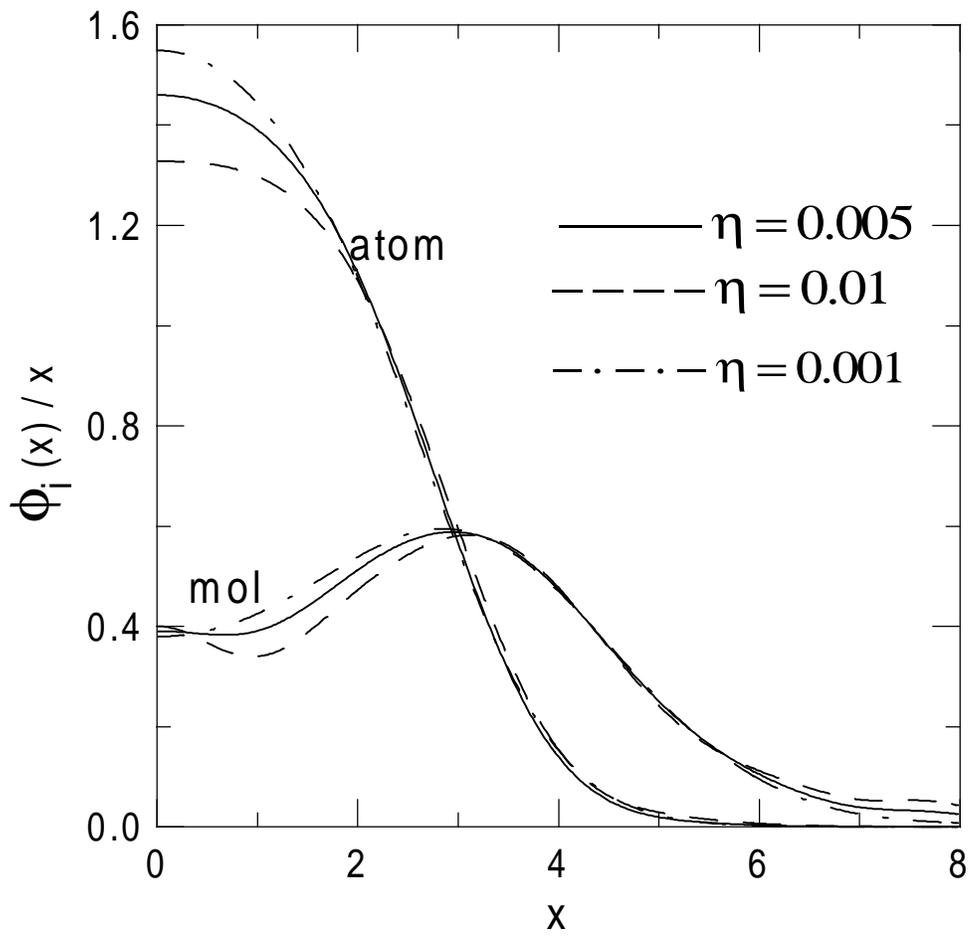
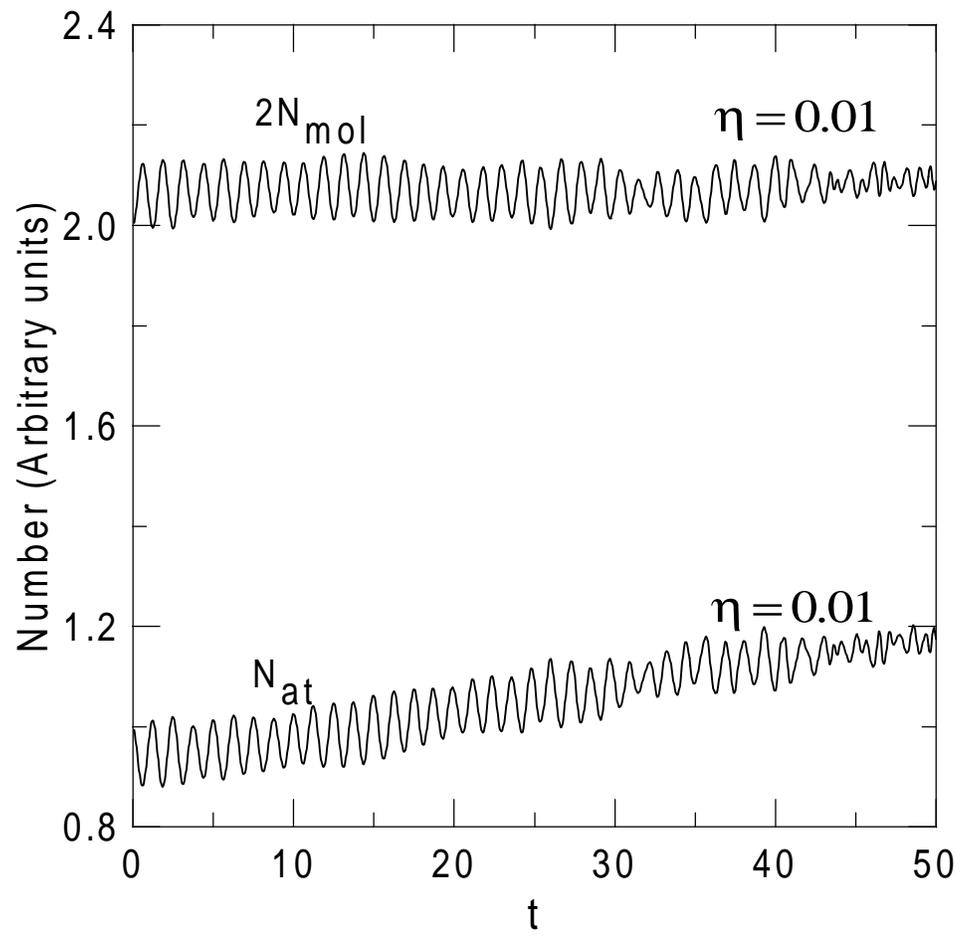
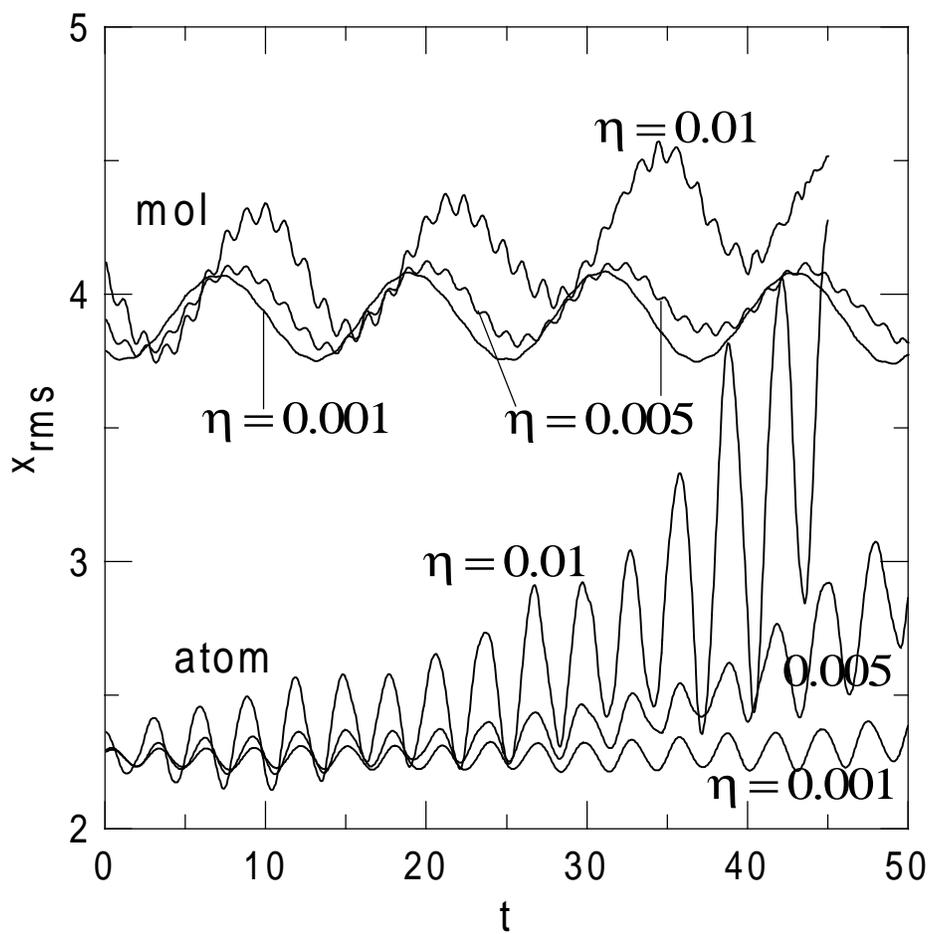
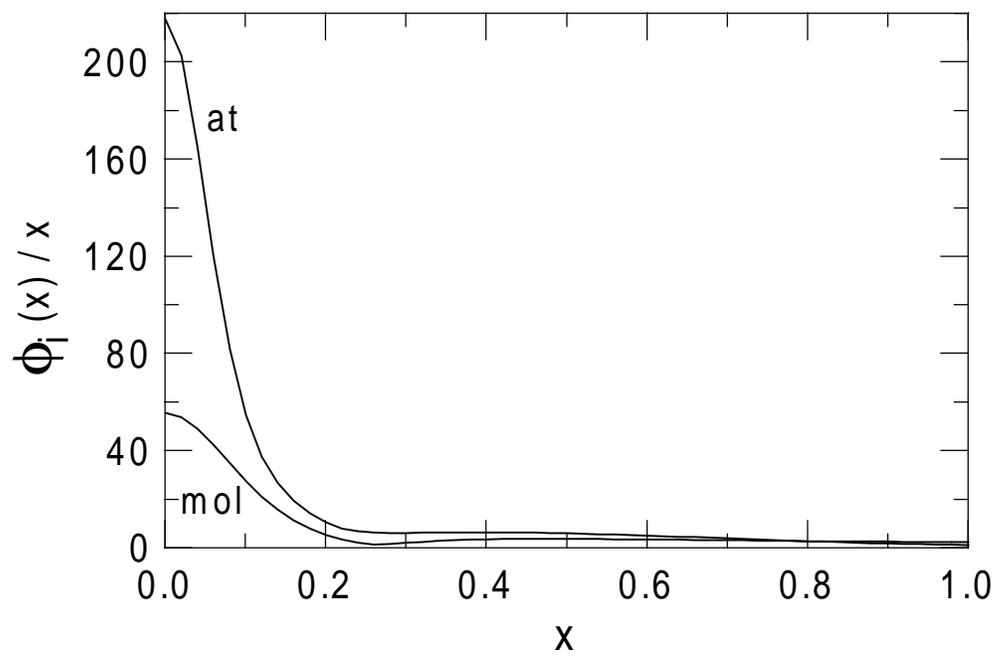
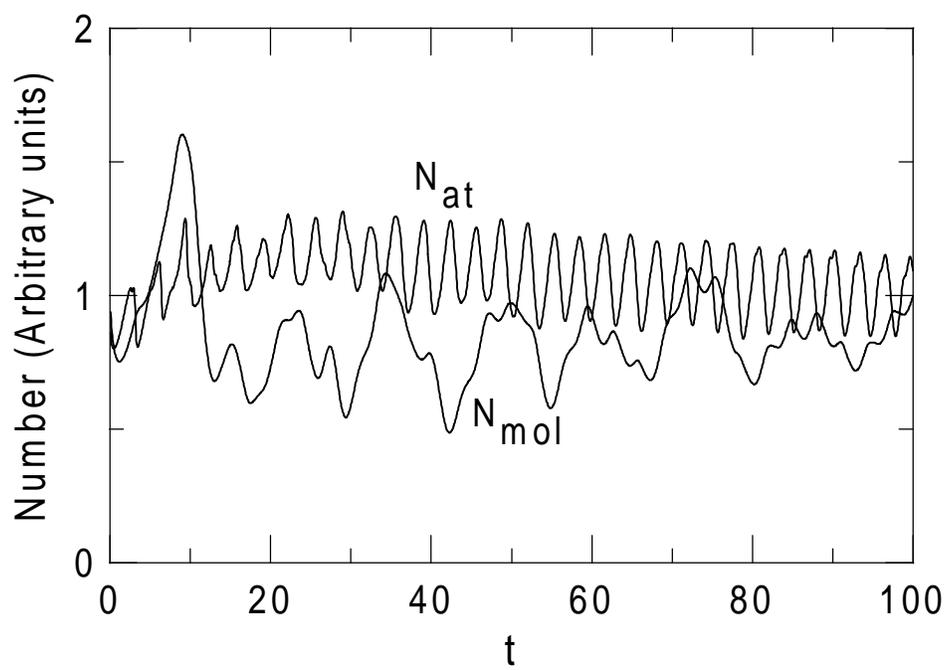
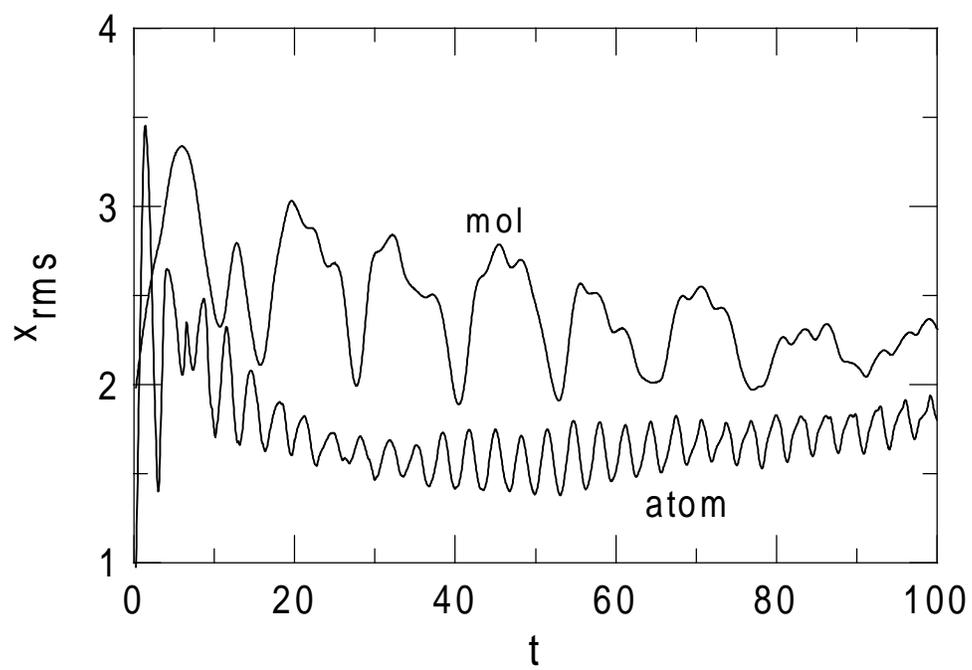
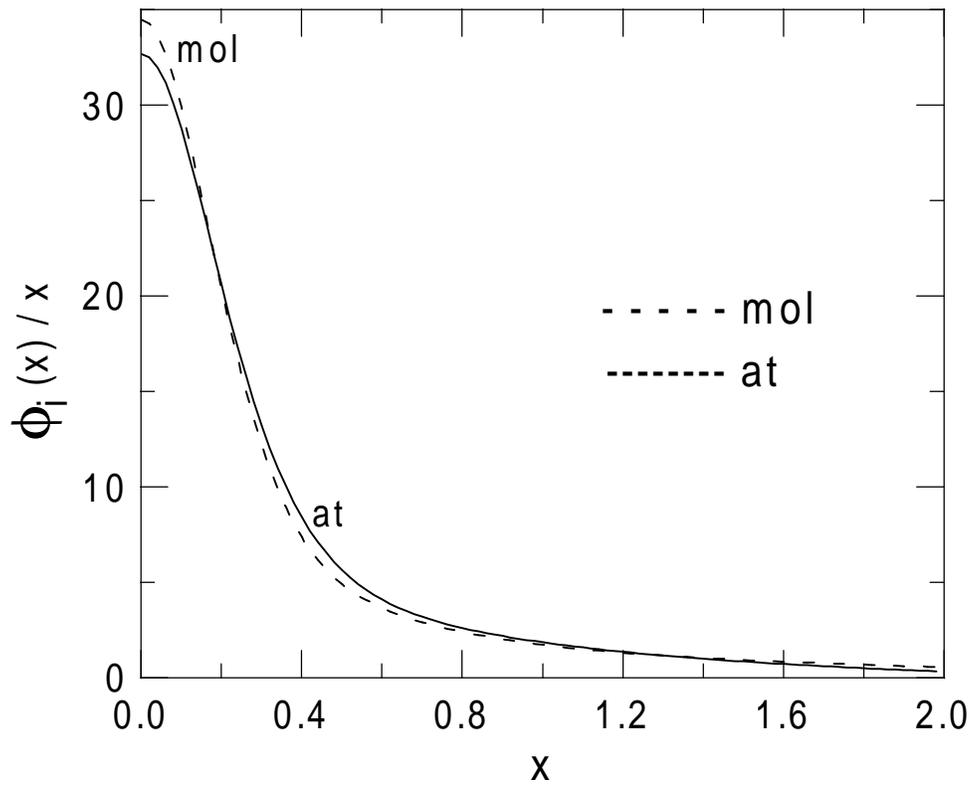
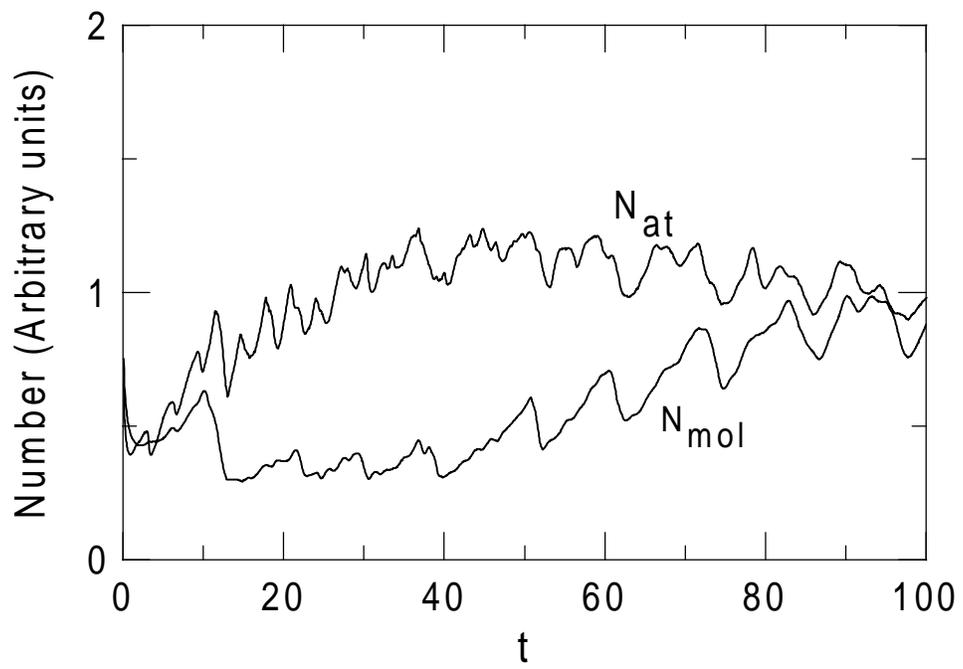
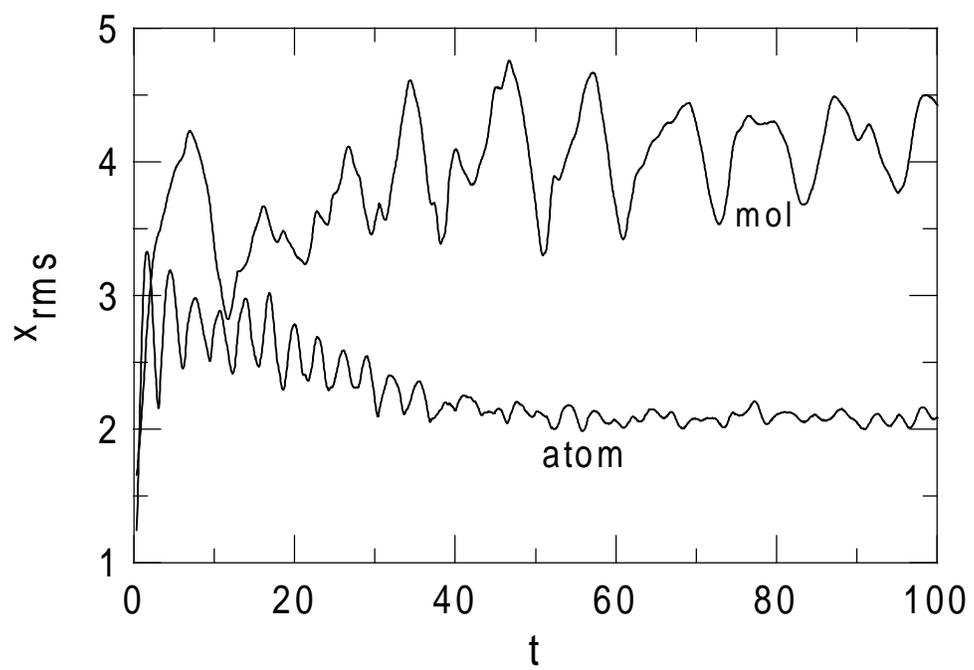
\end{document}